\documentclass[a4paper,12pt]{article}
\pdfoutput=1
\usepackage{amssymb,amsmath,bm}
\usepackage{a4wide}
\usepackage{color}
\usepackage{slashed}
\usepackage{cite}
\usepackage{hyperref}
\usepackage{units}
\usepackage{relsize}
\usepackage[latin1]{inputenc}
\usepackage{amsfonts}
\usepackage{lscape}
\usepackage{amsthm}
\usepackage{booktabs}
\usepackage{array}
\usepackage{rotating}
\usepackage{float}
\usepackage{multirow}
\usepackage{graphicx,rotating,amsmath,multirow,xcolor,colortbl,xcolor}
\usepackage{hyperref,pdflscape,slashed}
\definecolor{rosso}{cmyk}{0,1,1,0.4}
\definecolor{rossos}{cmyk}{0,1,1,0.55}
\definecolor{rossoc}{cmyk}{0,1,1,0.2}
\definecolor{blu}{cmyk}{1,1,0,0.3}
\definecolor{blus}{cmyk}{1,1,0,0.6}
\definecolor{bluc}{cmyk}{1,1,0,0.1}
\definecolor{verde}{cmyk}{0.92,0,0.59,0.25}
\definecolor{verdec}{cmyk}{0.92,0,0.59,0.15}
\definecolor{verdes}{cmyk}{0.92,0,0.59,0.4}
\definecolor{Gray}{gray}{0.95}

\font\tenrsfs=rsfs10 at 12pt
\font\sevenrsfs=rsfs7
\font\fiversfs=rsfs5
\newfam\rsfsfam
\textfont\rsfsfam=\tenrsfs
\scriptfont\rsfsfam=\sevenrsfs
\scriptscriptfont\rsfsfam=\fiversfs
\def\mathscr#1{{\fam\rsfsfam\relax#1}}

\usepackage[small,bf]{caption}
\hypersetup{colorlinks,bookmarksopen,bookmarksnumbered,
linkcolor=blus,pdfstartview=FitH,urlcolor=rossos,citecolor=verde}

\newcommand{\be}{\begin{equation}}
\newcommand{\ee}{\end{equation}}
\newcommand{\bea}{\begin{eqnarray}}
\newcommand{\eea}{\end{eqnarray}}
\newcommand{\beq}{\begin{equation}}
\newcommand{\eeq}{\end{equation}}
\newcommand{\beqa}{\begin{eqnarray}}
\newcommand{\eeqa}{\end{eqnarray}}

\counterwithin*{equation}{section}

\begin{document}


\begin{center}
\boldmath

{\textbf{\large Remembering Steven Weinberg}}

\unboldmath

\bigskip

\vspace{0.4 truecm}

{\bf Gian Francesco Giudice}
 \\[5mm]

{\it CERN, Theoretical Physics Department, Geneva, Switzerland}\\[2mm]

\end{center}
 
 \vspace{0.5cm}
 
\noindent {\it Contribution to the volume ``In Memory of Steven Weinberg" to appear in Nuclear Physics~B.}

\vspace{1cm}

Meeting Steven Weinberg has been one of the highest privileges of my professional life. In 1990, I was hired as a postdoc in his research group and spent two years in Austin, Texas. Those were extraordinarily stimulating years. Weinberg's group was relatively small, but comprised some of the brightest minds in physics of the time: Joe Polchinski, Willy Fischler, Vadim Kaplunovsky, Philip Candelas, among others. Even without Weinberg, this would have been one of the strongest groups in theoretical particle physics in the world. With Weinberg, it was nirvana for a young researcher like me.

Steve was the greatest hero in physics I could ever imagine. True, I have never had the opportunity of meeting Feynman, who may have been more impressive to a young researcher with his captivating enthusiasm and overwhelming brilliance. Steve was not the kind who overpowers you with words. During conversations he remained reflective and pensive, but anything that came out of his mouth was pure wisdom, expressed with a lucidity and logical rigor to leave you breathless. 

He liked to work alone and not to be constantly disturbed. To enter his office one had to go through the office of his loyal secretary Adele, who was protecting his privacy as a mastiff guards over its master's property. Steve was not spending everyday's life with common mortals, such as postdocs, but he wasn't ignoring them either. The encounters followed a recurrent pattern and, although relatively rare, were always meaningful and memorable. 

Once a week, there was the brown bag seminar, when Weinberg's group got together for lunch. Every week, a different member of the group was presenting some work in progress or an interesting paper they read, while the others were chewing their sandwiches. Speaking in front of so many eminent theorists and addressing their sharp questions was an experience one couldn't forget. Whenever Steve started to utter a question, I was holding my breath. He was often interrupting the speaker, not to prove them wrong but to understand the subject in his own way.

Another regular occasion for encounters were the lunches at the UT Faculty Club, which happened whenever an external speaker came to Austin to give the main seminar. All professors and postdocs of Weinberg's group were invited, together with the guest speaker. The Faculty Club was a few blocks away from the Physics Department, an easy and pleasant walk across campus. Nevertheless Steve was preferring to take his car, kindly inviting people for the ride. Almost everyone was declining, with the excuse that the weather was favorable for a short stroll (even when the outside temperature was above 100 Fahrenheit). I later understood that the true reason was Weinberg's poor driving skills (nobody is perfect), as his attention was easily diverted away from the traffic. I was one of the few who always accepted his invitations. Maybe it was just because, as an Italian, I have low standards for driving, but for me listening to Steve's conversation for five minutes was well worth the risk of a car accident. 

At my first visit to the Faculty Club, I was instructed about the rigid lunch etiquette by my friend and office mate Ramy Brunstein, a fellow postdoc more experienced than myself. Rule number one: there should be a single conversation at the table and you should never start small talk with the person sitting next to you. Rule number two: nobody (other than Weinberg) was allowed to change the topic of conversation. It was delightful to listen to Steve speaking
about physics in an informal setting. I will never forget the lunch in which our guest, a well-known phenomenologist, was explaining how subtle is the art of building models for new physics. Referring to Steve and himself collectively, he said: ``Because {\it we} model-builders..." Steve promptly interrupted him. ``I don't consider myself a model-builder," he said with an ironic smile on his face. ``In my life, I have built only {\it one} model."

Physics was not the only topic of conversation at Faculty Club lunches. Steve was a history connoisseur and enjoyed recounting episodes about the Civil War or making remarks about military strategy in famous battles of the Middle Ages. He also had a series of funny anecdotes about history. Once, after smiling at me as a way of asking for permission, he said that he wanted to tell a true story that happened in Italy, soon after the Racial Laws were introduced in 1938. The {\it podest\`a} (mayor) of a remote village in a forlorn valley in central Sicily was a scrupulous fascist, eager to show his loyalty to Mussolini. After receiving the dispatch listing the racial restrictions to be implemented in every town in Italy, he quickly telegraphed the {\it Duce}: ``Racial campaign ready. Please send Jews".

 Alas, after two years of postdoc, I started noticing that his repertoire was rather limited and that he kept repeating the same jokes to different guests. Experience with Faculty Club lunches helped me to realize that explaining complex concepts in physics came more natural to him than telling jokes. 

Seminars too followed a well-established ritual. Weinberg would enter the seminar room last and sit in the front row, with a pile of mail that was just handed to him by the devoted Adele. Soon after the beginning of the seminar, Steve would start asking questions. He was never overtly aggressive with speakers but he always wanted to get to the bottom of every issue. If something wasn't clear to him, he would stop the speaker and ask for explanations until he had fully understood the point. Eventually, if the answers didn't satisfy him, he would simply ignore the seminar and start opening noisily his mail (I mean tearing envelopes, unfolding letters and reading them; a concept of mail that young people today may not fully comprehend). Disinterest was his way of expressing dissent and humiliating the speakers.

Those were the years of the SSC catastrophe. Steve had become a vocal advocate of the project. He was regularly interviewed by journalists and often went to Washington to lobby for the SSC or participate in congressional hearings. Together with other postdocs, we were watching the interviews on TV. Mostly we were having fun laughing at Steve's difficulties with the ear-buds, which kept on falling from his ears during the interview. When back from trips to Washington, Steve liked to share with us his impressions of various senators and offer his opinions about the process. For me it was a thrilling experience: it felt like learning about the unfolding events directly from the man who was shaping the History of Physics. At the time, Steve seemed generally optimistic about the outcome of the SSC decision process in Washington.

I left Austin in the Fall of 1992. About one year later the SSC was cancelled. Weinberg mourned the decision and later said: ``A society that decides it will only support applied science and not waste money on pure science is likely to wind up with neither."
In the meanwhile, I had arrived at CERN, where LEP was in full swing, the LHC construction would be approved in December 1994, and LEP2 would start its gradual energy increase in 1996. I left the US to go to Europe just in time to participate in a new era of high-energy physics at CERN.

I remained in touch with Steve. I wouldn't call it a close friendship, but we exchanged email messages from time to time. Sometimes it was about physics: he liked the idea of Split Supersymmetry, and was intrigued by the method of calculating supersymmetry-breaking effects from wave-function renormalization and by anomaly mediation (work I did in collaboration with Nima Arkani-Hamed, Savas Dimopoulos, Markus Luty, Hitoshi Murayama, Riccardo Rattazzi, Andrea Romanino). Sometimes he was writing just to ask for news from CERN. We met in several occasions when he travelled to Europe. He once came to Padua, my hometown, and we spent some time together, also with his wife Louise. Faithful to his principles, he refused to enter religious buildings, let alone the synagogue, so missing a visit, which I had organized for him through some friends, to see a small Paduan cemetery hidden in a charming garden among old houses, where the 16th century Talmudist Meir Katzenellenbogen is buried.

Weinberg had a profound influence on me as a physicist. Because of the limited amount of time I spent with him, the influence came mostly from his writings. What I admire in those papers are his lucid thinking, logical clarity and brilliant insight. Few theorists match his vision of physics and foresight to anticipate new research directions. Take the example of effective field theories, which today are part of the toolkit of every particle physicist. Many of the founding ideas on effective theories came from his 1979 paper on {\it Phenomenological Lagrangians}.

Just as happened with his 1967 article (namely the paper that contains the {\it only} model Steve has ever proposed and that was later rewarded with a Nobel prize), also the message about effective field theories took a while to percolate through the physics community. I remember that, when I was a PhD student in the late 80s, some well-known authors were still claiming that renormalizability is a necessary requirement for any realistic quantum field theory. Luckily my PhD advisor Riccardo Barbieri, one of the greatest theorists of his time, was well aware of Wilson's and Weinberg's lessons and could pass them to me. 

Today the database Inspire lists 46 papers by Weinberg with more than 500 citations, and 78\% of them are single-authored. Each of them is a masterpiece that broke new ground in a different subject, testifying to the amazing breath and importance of Weinberg's contribution to physics. Any theorist would consider a supreme accomplishment of their career to have written only one of them. Weinberg wrote 46 such papers. Every student in theoretical particle physics should read all of them at least once, not so much to learn about their content, which by now is standard material that can be found in any textbook on the subject, but to learn what it means to do research in theoretical physics.

Not only his technical papers, but his books too shaped my views on physics. His monumental textbooks, from gravitation and cosmology to quantum field theory, are still an essential part of any theoretical physicist's education. I am particularly fond of his books addressed to the general public, which contain powerful lessons on the meaning of science and its role in society. Reading his essays was crucial for me to form my own perspective on these issues. Before giving public talks, I often refresh my memory by reading once again some of the chapters of his popular books. Besides always being a pleasure, it is also a way for me to return to the old master and listen to familiar advice.

Weinberg is a hero for all of us who had the privilege of meeting him and a lasting intellectual beacon for future generations.

\end{document}